# Condensed matter realization of fermion quasiparticles in Minkowski space

Xiao Dong[1,†], QuanSheng Wu[2,3,†], Oleg V. Yazyev[2,3], Xin-Ling He[1], Yongjun Tian[4], Xiang-Feng Zhou[1,4*] and Hui-Tian Wang[1,5*]

[1]*Key Laboratory of Weak-Light Nonlinear Photonics, Ministry of Education, School of Physics, Nankai University, Tianjin 300071, China*
[2]*Institute of Physics, École Polytechnique Fédérale de Lausanne (EPFL), CH-1015 Lausanne, Switzerland*
[3]*National Centre for Computational Design and Discovery of Novel Materials MARVEL, Ecole Polytechnique Fédérale de Lausanne (EPFL), CH-1015 Lausanne, Switzerland*
[4]*State Key Laboratory of Metastable Materials Science and Technology, Yanshan University, Qinhuangdao 066004, China*
[5]*National Laboratory of Solid State Microstructures and Collaborative Innovation Center of Advanced Microstructures, Nanjing University, Nanjing 210093, China*
[†]These authors contributed equally to this work

## Abstract

"What is the difference between space and time?" is an ancient question that remains a matter of intense debate.[1] In Newtonian mechanics time is absolute, while in Einstein's theory of relativity time and space combine into Minkowski spacetime. Here, we firstly propose Minkowski fermions in 2+1 dimensional Minkowski spacetime which have two space-like and one time-like momentum axes. These quasiparticles can be further classified as Klein-Gordon fermions and Dirac-Minkowski fermions according to the linearly and quadratically dispersing excitations. Realization of Dirac-Minkowski quasiparticles requires systems with particular topological nodal-line band degeneracies, such as hyperbolic nodal lines or coplanar band crossings. With the help of first-principles calculations we find that novel massless Dirac-Minkowski fermions are realized in a metastable bulk boron allotrope, *Pnnm*-B$_{16}$.

In crystalline solids the macroscopic transport phenomena can be described by resorting to the semi-classical approximation that considers dynamics of wave packets constructed from Bloch wave functions[2]. These wave packets, behaving like distinct particles, are commonly referred to as quasiparticles[3] or fermions. The band structures of crystalline solids describe the relationship between the energy and momentum of such quasiparticles. The extrema of the valence conduction bands in a classical semiconductor are described by the band dispersion relationship $E = \frac{p_x^2}{2m_{xx}} + \frac{p_y^2}{2m_{yy}} + \frac{p_z^2}{2m_{zz}}$,[4] where $E$ is the energy of the quasiparticle, $p_x$, $p_y$, $p_z$ represent the components of momentum proportional to the wave vector, $\vec{p} = \hbar \vec{k}$, and the effective masses $m_{xx}$, $m_{yy}$ and $m_{zz}$ are defined in the frame of principal axes reflecting the anisotropy of the band dispersion. The case of $m_{xx}$, $m_{yy}$, $m_{zz} > 0$ describes the conduction band minimum and the corresponding charge carriers are referred to as the "electron quasiparticles". Similarly, the case of $-m_{xx}$, $-m_{yy}$, $-m_{zz} < 0$ corresponds to the "hole quasiparticles" at the top of the valence band. Our story, however, will consider scenarios described by effective masses $m_{xx}$, $m_{zz} > 0$, $-m_{yy} < 0$ and $-m_{xx}$, $-m_{zz} < 0$, $m_{yy} > 0$.

Here, we employ the metric theory for describing the qusiparticles. In the nonrelativistic case, we impose the classical momentum-energy relationship, $E = p^2/2m$, and define $p^2 = \eta_{\mu\nu} p^\mu p^\nu$, where $\mu, \nu = x, y, z$ and $\eta_{\mu\nu}$ is the spatial metric. The introduction of $\eta_{\mu\nu}$ implies that the fermion's properties are determined by the ground-state atomic structure and the effects of periodic potential field. In a general case, $\eta_{\mu\nu}$ in its principal axes system is $\eta_{xx} = \frac{m}{m_{xx}} \neq \eta_{yy} = \frac{m}{m_{yy}} \neq \eta_{zz} = \frac{m}{m_{zz}}$, and its off-diagonal elements are zero. We now assume $\eta_{yy} = -\frac{m}{m_{yy}} < 0$, $\eta_{xx} = \frac{m}{m_{xx}} > 0$ and $\eta_{zz} = \frac{m}{m_{zz}} > 0$. It implies that $\eta_{xx}, \eta_{yy}, \eta_{zz}$ not only have different values but also different signs, and thus change the space from ellipsoidal Euclid space to the hyperbolical Minkowski space[5]. We define Minkowski fermions as the quasiparticles with Minkowski spatial metric.[6]

By employing Lorentz transformation, Einstein used Minkowski space to describe spacetime in the theory of relativity. The Minkowski metric can be written as

$$\eta = \begin{pmatrix} \frac{m}{m_{xx}} & 0 & 0 \\ 0 & -\frac{m}{m_{yy}} & 0 \\ 0 & 0 & \frac{m}{m_{zz}} \end{pmatrix}. \tag{1}$$

Below, we will refer to the $x$ and $z$ axes as the space-like axes, while to the $y$ axis as the time-like axis.

As follows from the Minkowski metric, Minkowski fermions are the quasiparticles at the saddle point of an electronic band. Physical phenomena occurring at the saddle points have been extensively discussed in past, e.g. in the context of Lifshitz transition[7,8], and the Landau level transition from continuous to discretized spectrum under different orientations of the magnetic field[9,10]. The saddle-point singularity is in fact very common at the high-symmetry momenta of the energy band structures from simple materials such as alkali metal lithium (see Supplementary Fig. S1) to a flat $CuO_2$ planes in high-temperature superconductor $YBa_2Cu_3O_7$ [11]. However, all of them are topological trivial due to the lack of band crossing.

We will now make a step forward and extend the one-band Minkowski fermions model to the 2-band model by introducing the electron-hole symmetry. The Hamiltionian becomes

$$\hat{H} = \pm(\frac{\hat{p}_x^2}{2m_{xx}} - \frac{\hat{p}_y^2}{2m_{yy}} + \frac{\hat{p}_z^2}{2m_{zz}}), \tag{2}$$

where $\hat{p}_\mu = -i\hbar \partial/\partial\mu$ and the symbol $\pm$ distinguishes electrons from holes. In this equation, $y$ is the time-like axis, hence we set $y = ct$ (note, time $t$ is just a mathematical notation by analogy and should be distinguished with the true time). By introducing normalization $x' = x\sqrt{2m/m_{xx}}, t' = t\sqrt{2m/m_{yy}}, z' = z\sqrt{2m/m_{zz}}$ and natural units, $\hbar = c = 1$, Hamiltonian (2) becomes

$$\hat{H} = \mp\frac{\hbar^2}{m}(\frac{\partial^2}{\partial x'^2} - \frac{\partial^2}{c^2\partial t'^2} + \frac{\partial^2}{\partial z'^2}) = \mp(-\partial_t^2 + \partial_x^2 + \partial_z^2)/m. \tag{3}$$

Equation (3) is equivalent to free 2+1 dimensional Klein-Gordon equation[12] $-\partial_t^2 \psi + \nabla^2 \psi = m^2 \psi$, in (x,z,t) spacetime.

Klein-Gordon equation describes spinless bosons, such as the pion and the recently discovered Higgs boson. However, here we build Klein-Gordon-like "fermions" in 2+1 dimensional (2+1D) Minkowski spacetime. This implies one axis is chosen as time-like dimension and the 3 dimensional (3D) Euclid space is mapped onto the 2+1D Minkowski spacetime of the quasiparticles. Electrons and holes in real space can now be treated as Klein-Gordon fermions, a new type of fermions that should be distinguished from the massless Dirac fermions and the massive fermions associated with the quadratic band dispersion.

For the band dispersion related to the Klein-Gordon fermions, the valence and conduction bands touch at the Fermi level, which is set to zero energy for simplicity (Figs. 1 a,b). The two nodal lines, composed of crossing points of the valence and conduction bands, coplanar cross in the $p_x$-$p_y$ plane. That is, Minkowski fermions can be found in systems with two coplanar crossing nodal lines[13].

The nodal-chain metals hosting a new type of fermion quasiparticles was proposed to be realized in IrF$_4$[14] and TiB$_2$[15]. Excitations in these systems can be treated as an extension of Klein-Gordon fermions because the energy-momentum relationship was $E = \pm((p_x - b)^2/m_{xx} - p_y^2/m_{yy} + p_z^2/m_{zz} - b^2) + O(p_y p_z)$, where a second order perturbation term $O(p_y p_z)$ is attributed to band repulsion resulting in the formation of chain instead of uniparted hyperboloid.

The energy-momentum relationship underlying Klein-Gordon fermions (3) is still a quadratic equation. We can now follow the steps of Paul Dirac[16] to build a linear equation with relativistic theory in Minkowski spacetime. In a 2+1D system in relativistic theory

$$E = e(p_t c - \sqrt{p_x^2 c^2 + p_z^2 c^2 + m^2 c^4}), \qquad (4)$$

where $e = \pm 1$ identifies the pair of fermions and their corresponding antiparticles, i.e. electrons ($e = -1$) and holes ($e = 1$). Then, we introduce the mass anisotropy obtaining

$$E = e(p_y c_t - \sigma_x p_x c_x - \sigma_y p_z c_z - \sigma_z m c_t^2),  \tag{5}$$

where $\sigma_x, \sigma_y, \sigma_z$ are the 2×2 Pauli matrixes. Here, $c_x, c_t, c_z$ mean that the velocity of light is a vector instead of a constant in an anisotropic periodic field.

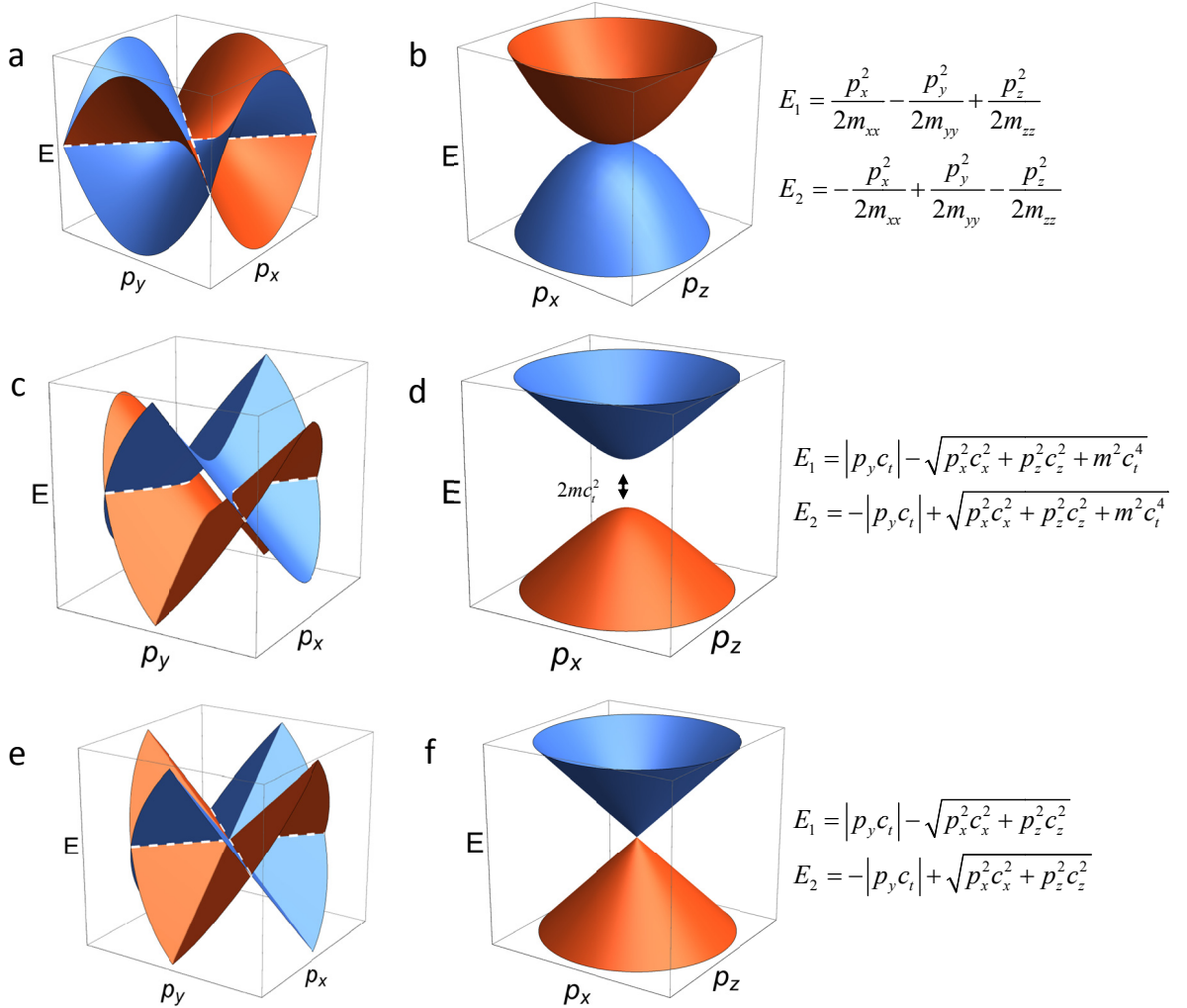

Fig. 1 Projections of band dispersions of the proposed three types of Minkowski fermions: Klein-Gordon fermions in (a) the $p_x$-$p_y$ plane and (b) the $p_x$-$p_z$ plane; massive Dirac-Minkowski fermions in (c) the $p_x$-$p_y$ plane and (d) the $p_x$-$p_z$ plane; massless Dirac-Minkowski fermions in (e) the $p_x$-$p_y$ plane and (f) the $p_x$-$p_z$ plane. The black dashed line shows the nodal line at the intersection of the valence band and the conduction band. The $p_x$-$p_y$ and $p_x$-$p_z$ plane projections are equivalent.

By solving equation (5), we obtain $E$ as a group of four energy bands:

$$E = e(p_y c_t \pm \sqrt{p_x^2 c_x^2 + p_z^2 c_z^2 + m^2 c_t^4}) = \begin{pmatrix} p_y c_t + \sqrt{p_x^2 c_x^2 + p_z^2 c_z^2 + m^2 c_t^4} \\ p_y c_t - \sqrt{p_x^2 c_x^2 + p_z^2 c_z^2 + m^2 c_t^4} \\ -p_y c_t + \sqrt{p_x^2 c_x^2 + p_z^2 c_z^2 + m^2 c_t^4} \\ -p_y c_t - \sqrt{p_x^2 c_x^2 + p_z^2 c_z^2 + m^2 c_t^4} \end{pmatrix}, \quad (6)$$

so the Dirac-Minkowski fermions are well defined in a crystal with local dispersion relation from Eq. (6).

Compared to the Minkowski fermion model considered in the beginning, the use of Dirac undetermined coefficient method to decrease the equation from quadratic to linear and the introduction of the Pauli matrices makes Dirac-Minkowski fermions to be a group of four fermions. On the other hand, compared to the 2D Dirac fermions, like graphene, the introduction of the new time-like axis, the $ep_y c_t$ term, breaks the original symmetry, resulting in two bands being split into four non-degenerated bands.

Here, if we only consider the conduction band and the valence band near the Fermi level, i.e. the middle two bands, the dispersion

$$E = \begin{pmatrix} |p_y c_t| - \sqrt{p_x^2 c_x^2 + p_z^2 c_z^2 + m^2 c_t^4} \\ \sqrt{p_x^2 c_x^2 + p_z^2 c_z^2 + m^2 c_t^4} - |p_y c_t| \end{pmatrix} \quad (7)$$

describes the energy-moment relation of electrons and holes. Since $|p_y c_t|$ and $\sqrt{p_x^2 c_x^2 + p_z^2 c_z^2 + m^2 c_t^4}$ have different signs, it satisfies our definition of Minkowski fermions.

The Dirac-Minkowski fermions can be further divided into two classes according to their mass. As shown in Fig. 1d, for massive Dirac-Minkowski fermions with $m \neq 0$, a gap of $2mc_t^2$ is opened at $\vec{p} = 0$ with a dispersion of normal semiconductor in the $p_x$-$p_z$ plane. However, in the $p_x$-$p_y$ plane, conduction and valence bands touch forming the hyperbolic nodal lines at $E = 0$.

For $m = 0$ four bands form a degeneracy at $\vec{p} = 0$ resulting in an eight-component fermions when spin degeneracy is taken into account (Fig. 1e and f). In the $p_x$-$p_z$ plane, it has a two-dimensional (2D) Dirac cone band dispersion, while in the $p_x$-$p_y$ plane, the conduction and valence bands are joined together with two coplanar crossing nodal lines (Fig. 1e), similar to the case of Klein-Gordon fermions.

In the case of Klein-Gordon fermions, the energy-momentum relationship is quadratic in all three dimensions. For massive Dirac-Minkowski fermions, the energies show linear dispersion along the time-like momentum direction and quadratic dependence on the space-like momentum. For the massless Dirac-Minkowski fermions, the energies have linear dispersion along all three momentum axes, no matter whether they are time-like or space-like. Different from normal quasiparticles (free-electron-like, Dirac or Weyl fermions) in Euclid space, all the Minkowski fermions have one time-like axis, which has different effective masses with opposite signs, compared to the other two space-like axes. Minkowski systems are semimetals with nodal lines because opposite signs of effective masses lead to the degeneracies of the valence and conduction band. Massive Dirac-Minkowski fermions are characterized by hyperbolic nodal lines while Klein-Gordon and massless Dirac-Minkowski fermions have coplanar crossing nodal-line degeneracies. Hyperbolic nodal lines and coplanar crossing nodal lines are different from previously discussed nodal line forms, such as elliptical nodal lines[17].

Furthermore, the first derivative of energy, $\partial E/\partial \vec{p}$ is not continuous at the nodes. This implies that the derivation chain rule is violated and the principal axis approximation may not hold. For all the three types of Minkowski fermions (Klein-Gordon, massive Dirac-Minkowski and massless Dirac-Minkowski fermions), one cannot simply use three principal axes to represent the behavior of electrons or holes. In other words, one cannot guess the entire energy-momentum relation of fermions by only knowing three orthogonal one-dimensional dispersions. For example, for the band structure of Minkowski fermions, $E = \pm(\frac{p_x^2}{2m_{xx}} - \frac{p_y^2}{2m_{yy}} + \frac{p_z^2}{2m_{zz}})$ shown in Fig. 1, its $E(p_x)|_{p_y,p_z}, E(p_y)|_{p_x,p_z}, E(p_z)|_{p_x,p_y}$ dispersions are the same as for a classical Euclid semimetal $E = \pm(\frac{p_x^2}{2m_{xx}} + \frac{p_y^2}{2m_{yy}} + \frac{p_z^2}{2m_{zz}})$. This means that 2D and even 3D dispersions are necessary to evaluate a fermion behavior, especially for Minkowski fermions.

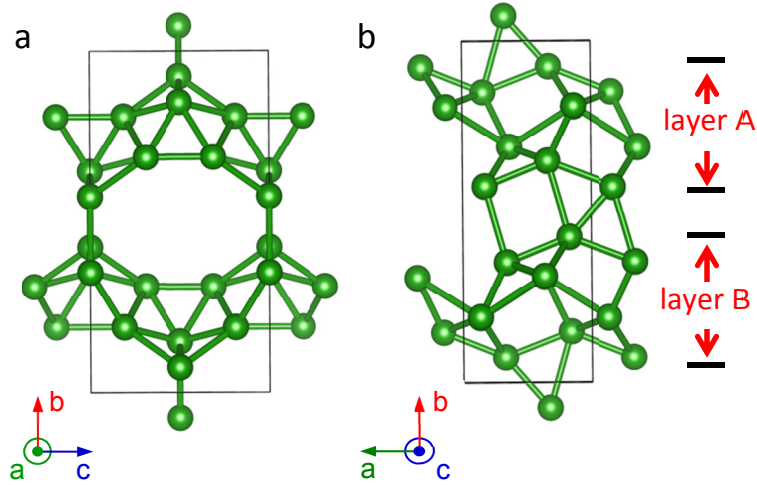

Fig. 2 The crystal structure of *Pnnm*-B$_{16}$ in (a) [100] and (b) [001] directions. B atoms are shown green.

Given the ability of spin-orbit interaction to lift the nodal-line degeneracy, we will discuss the materials based on light elements belonging to the second row of the periodic table in which these relativistic effects are intrinsically weak. Similar to carbon, elemental boron can form a series of metastable phases in zero, one, two or three dimensions, but the bonding scenarios in boron system is more complex due to peculiar multicenter bonds. The presence of band degeneracies is determined by the combination of bonding symmetry and lattice symmetry. Therefore, we expect finding new types of fermions in metastable semimetallic boron materials by varying bonding and lattice symmetry.

The search for a metastable boron material hosting Minkowski fermions was performed allowing up to 32 atoms per primitive cell at zero pressure, using the evolutionary structure prediction algorithm[18] as implemented in the USPEX code[19]. Metastable phases with 60 lowest energies were kept for subsequent electronic structure calculations in order to identify the semimetal band structures and special nodal line degeneracies. One of the investigated allotropes of boron, named *Pnnm*-B$_{16}$, showed the Dirac-Minkowski fermions near the Fermi level.

*Pnnm*-B$_{16}$ contains 16 atoms per primitive cell (Figs. 2a and 2b). At zero pressure, the lattice parameters are $a$ = 3.17 Å, $b$ = 8.39 Å and $c$ = 4.46 Å. Boron atoms occupy the Wyckoff positions 8h (0.635, 0.810, 0.312), 4g (0.669, 0.148, 0) and 4g (0.172, 0.074, 0), respectively. This structure can be considered as the AB stacking of the previously investigated 2D structure, *Pmmn*-B$_8$, which was proved to have anisotropic Dirac fermions in 2D Euclid space.[20]. The position of layer A was shifted 0.17*a* (0.53 Å) from layer B along [010] direction and the interlayers are connected by multicenter B-B bonds. As a result, compared with *Pmmn*-B$_8$, *Pnnm*-B$_{16}$ inherit its Dirac fermion band dispersion in the $p_x$-$p_z$ plane, but also has a different dispersion in the $p_y$ direction that allows to assign space-like axes to *x* and *z* and the time-like axis to *y*.

As a metastable phase at atmospheric pressure, *Pnnm*-B$_{16}$ has a binding energy of -6.12 eV/atom relative to a single B atom, while the binding energies are -6.40 eV/atom of the stable phase α-boron, -6.01 eV/atom of single sheet of the α-phase and -6.06 eV/atom of 2D-*Pmmn*-B8. *Pnnm*-B$_{16}$ is more stable than many suggested structural phases of boron. In addition, the calculated phonon spectrum indicates that this phase is dynamically stable at ambient condition (supplementary Fig. S2).

As shown in the band structure in supplementary Fig. S3, *Pnnm*-B$_{16}$ is a semimetal with conduction and valence bands forming several degeneracies at the Fermi level. These degeneracies are protected by symmetry, that is the band crossings belong to different symmetry representations.[21] Particularly, in the Γ-X-S plane in the Brillouin zone (BZ) of *Pnnm*-B$_{16}$, the little point group is C$_s$, which has two irreducible representations, A' and A''. As shown in Fig. 3a, at point Γ, there are four energy levels with the sequence of A'', A', A', A'' while two A' (A'') levels on the X-S line in the BZ boundary are degenerate as required by symmetry. This guarantees the presence of two intersections of A'- and A''-symmetry bands between Γ and X and between Γ and S. As shown in Fig. 2a, one crossing is near the Fermi level in both Γ-X and Γ-S segments, which indicates the presence of a nodal line that crosses these segments. Unlike in common nodal-line semimetal, the double degenerate levels A' and A'' in the X-S line cross each other and the crossing at (0.188, 0.5, 0) is also protected by symmetry. Therefore, there is a quadruple degenerate (eight-component, if spin is considered) point in the X-S line and two nodal lines coplanar intersect at this point as shown in Fig. 2b. Dispersion of these four bands satisfies the quaternion of massless Dirac-Minkowski fermions and implies there is a Dirac-Minkowski point on the X-S line.

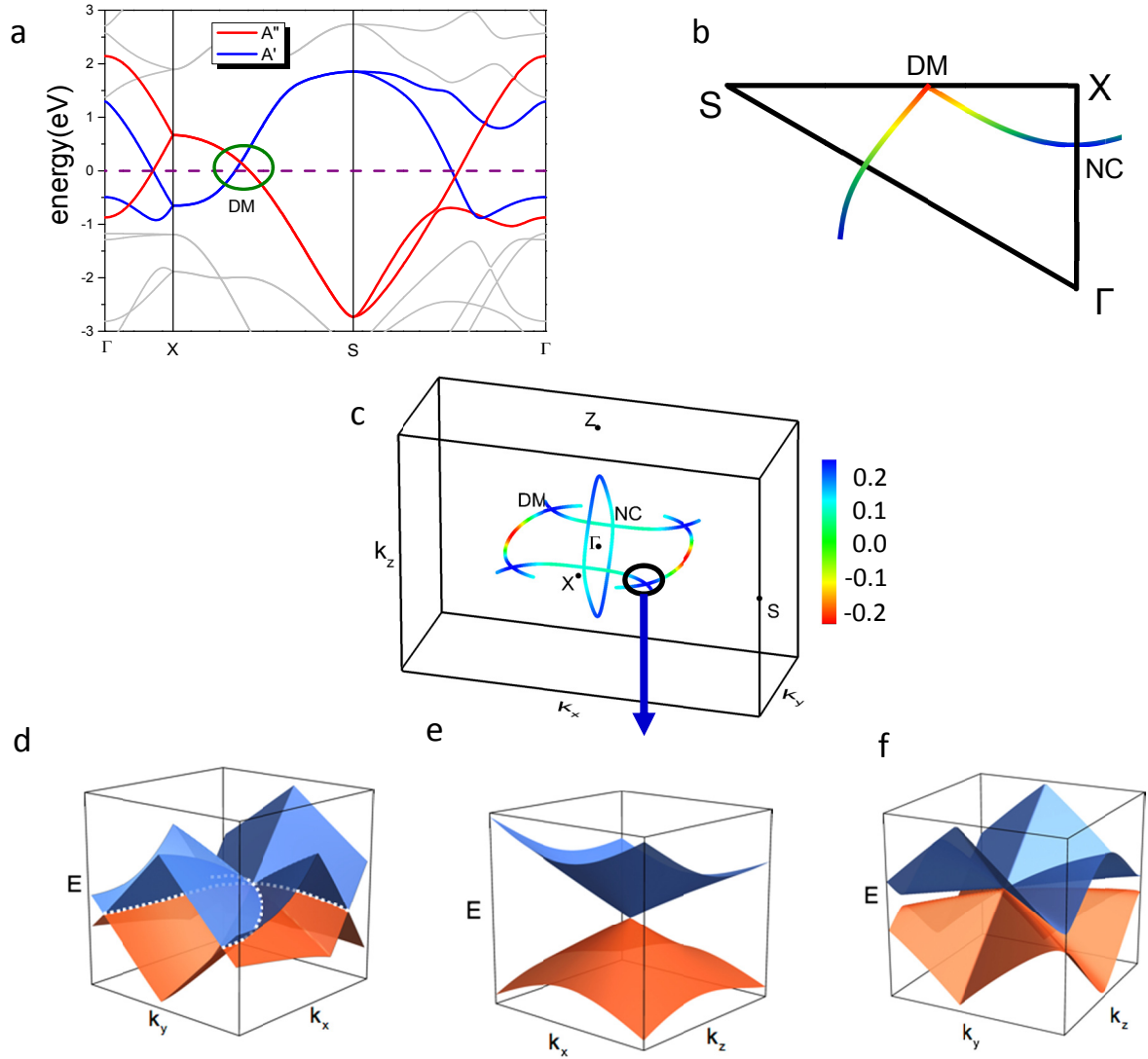

Fig. 3 (a) Band structure of *Pnnm*-B$_{16}$ along the Γ-X-S-Γ line. (b) Nodal lines cross Γ-X and Γ-S lines, respectively and join at the Dirac-Minkowski point. DM represents Dirac-Minkowski point and NC indicates the nodal chain points (c) The nodal lines in the entire Brillouin zone of *Pnnm*-B$_{16}$. The color shows the energy relative to the Fermi level.. (d)-(f) Projections of the energy dispersion at the Dirac-Minkowski point on (d) the $k_x$ - $k_y$ plane, (e) $k_x$ - $k_z$ plane and (f) $k_y$ - $k_z$ plane. The Fermi energy is set to zero.

As stated above, spin-orbit coupling can break the symmetry lifting the degeneracy at the Dirac-Minkowski point opening a gap. However, as discussed in supplementary Fig. S3b, the effect of spin-orbit coupling is quite small in *Pnnm*-B$_{16}$, opening a gap of only 1.35 meV. Therefore, in the following discussion we will ignore the effect of spin-orbit coupling in *Pnnm*-B$_{16}$.

The Dirac-Minkowski point is the intersection of the nodal line and the BZ boundary. Because this intersection with BZ boundary is not orthogonal, the nodal line reflects back into the first BZ resulting in a crossing of two nodal lines. In fact, such reflection at the BZ boundary is a simple and efficient way of identifying Dirac-Minkowski systems.

As shown in Fig. 3c, there are four Dirac-Minkowski points at the first Brillouin boundary protected by the time-reversal and mirror symmetries at ($\pm$0.0592, $\pm$0.0596, 0) × 2$\pi$ Å$^{-1}$, or ($\pm$0.188, $\pm$0.5, 0) in terms of the fractional coordinates. We consider (0.188, 0.5, 0) as a representative Dirac-Minkowski point and place it at the origin. Coefficients $c_x^+$ = 33.7 eV·Å, $c_x^-$ = 18.2 eV·Å, $c_t = c_y$ = 19.0 eV·Å, $c_z$ = 26.5 eV·Å and $m = 0$ were obtained by fitting Eq. (7) to the calculated Fermi velocities of 0.82×10$^6$, 0.44×10$^6$, 0.46×10$^6$ and 0.64 ×10$^6$ m/s for the $x^+$, $x^-$, $y$ and $z$ directions, respectively. The difference between the $x^+$ and $x^-$ Fermi velocities implies an asymmetric band dispersion relation in the $x$ direction, which originates from two bands with distinct wave functions, i.e. A' and A" bands shown in Fig. 3a. All Fermi velocities have in the discussed material are comparable to the one of graphene, 34 eV·Å (0.84×10$^6$ m/s) in our calculation and about 1.0×10$^6$ m/s in previous reports[22].

In the vicinity of the Dirac-Minkowski point, the two nodal lines can be considered straight lines, as imposed by Eq. (7) for $m = 0$. However, away from the Dirac-Minkowski point, the nodal lines show noticeable bending (Fig. 3b). This bending can be explained by the fact that the Dirac-Minkowski point is located away from Γ point. In the nearly-free-electron model and the $k\cdot p$ perturbation theory, the band energy tends to a form of $E(\vec{k}) \sim \hbar^2 k^2$ ($k$ is the vector relative to the Γ point) to minimize its kinetic energy if we ignore the periodic potential. In *Pnnm*-B$_{16}$, the nodal lines tend to bend into a ring with the center at the Γ point. Interestingly, if we define the middle of nodes for a fixed $k_y$ around Dirac-Minkowski point to make up a "geodesic line", this geodesic line will deviate true time-like $k_y$ axis (supplementary Fig. S4). If there is an observer in the fermion spacetime, he will conclude that the energy can warp the time and space, which is analogous to Einstein's general relativity.

As shown in Figs. 3d and 3e, in both $k_x$ - $k_y$ plane ($k_z = 0$) and $k_x$ - $k_z$ plane ($k_y = 0$), the band structure accords with the Dirac-Minkowski model. However, for $k_z \neq 0$ and $k_y \neq 0$ protection by symmetry is lost and a small gap opens as shown in Fig 3f. As a result, the band dispersion deviates from the Dirac-Minkowski model, even though this deviation is quite small. The most pronounced deviation in the $k_x$ - $k_z$ plane is along the line $k_y/k_z = \pm c_z/c_y$, where should be another nodal line in perfect Dirac-Minkowski model. On this line, the band repulsion gives a new Fermi velocity $v' = \partial E/\partial \vec{n} = \pm 1.3$ eV·Å with $\vec{n}$ being the line direction, although it is much smaller than $c_x^+$, $c_x^-$, $c_z$ and $c_t$. We suggest that it is a second order perturbation of $O(k_y k_z)$ and in many cases the band gap opening can be ignored when studying the electronic transport phenomena of the Dirac-Minkowski fermions.

Besides the nodal lines forming the Dirac-Minkowski points, there is another nodal line in the Γ-X-Z plane. These two types of nodal lines cross on the Γ-X line forming a nodal chain point. By time-reversal symmetry, there should be one pair of nodal chain points on both sides of the Γ point.

We find that *Pnnm*-B$_{16}$ has the coplanar nodal-line crossing hosting Dirac-Minkowski fermion quasiparticles. In fact, nodal-line semimetal phases are found in many materials including TiB$_2$[15], three-dimensional graphene networks[17], Cu$_3$PdN[23], etc. One can expect to find novel nodal line topologies and other realizations of massive or massless Dirac-Minkowski fermion quasiparticles in condensed systems. Furthermore, photonic crystals and artificial microstructures[24] host topological phases, but these systems can often be engineered in a controlled way, thus one may expect new types of Dirac-Minkowski quasiparticles to be discovered. Moreover, lattice field theory is an effective method to solve non-perturbative problem and our work suggests that a well-designed lattice can change from ground 3D and 4D Euclid space to 2+1 and 3+1D Minkowski spacetime, which implies a microscopic inhomogeneity and structure might lead a macroscopic change in space characteristics, i.e. a microscopic space axis change to a macroscopic time axis.

In 2005, Science journal listed 125 questions in science, one of which is "Why is time different from other dimensions?".[1] Although this question is still open in the case of real spacetime, we can give an insight in the view of condensed matter physics. Here we build fermions with 2 space and 1 time dimensions from a crystal in 3 purely "space" dimensions. So time axis can be considered as a derivative of space axis in the presence of a periodic field. In 1972, P. W. Anderson said famous quote "more is different",[25] and here we can say this difference can as large as that between time and space.

In conclusion, we built three types of fermions in 2+1 dimensional Minkowski spacetime: Klein-Gordon fermions, and derived from them, massive and massless Dirac-Minkowski fermions. These novel fermions have two space-like momentum axes and one time-like axis. Their condensed matter realization requires a semimetal with particular topology of nodal lines, such as hyperbolic curve or coplanar crossing, in contrast to the usual elliptical nodal lines. Here, using symmetry and reflection at the BZ boundary, we find Dirac-Minkowski fermions in a metastable boron phase, *Pnnm*-$B_{16}$. This proves that one can build time-like axis in presence of periodic field in a pure 3D space system. The prediction of the new type fermions in Minkowski space will change our comprehension of fermions and time, and have impact on our understanding of condensed matter physics and fundamental physics.

**METHODS**

Ab initio calculation were performed using density functional theory (DFT) within the Perdew-Burke-Ernzerhof (PBE) functional and local density approximation (LDA) exchange-correlation functional of Ceperley and Alder, as parameterized by Perdew and Zunger (CA-PZ)[26] in the framework of the all-electron projector augmented wave (PAW) method[27] as implemented in the VASP code[28]. For B atoms we used PAW potentials with 1.7 a.u. core radius and $2s2p$ electrons treated as valence. The evolutionary algorithm code USPEX[19], used here for predicting new stable structures, searches for the lowest-enthalpy structures at given pressure and is capable of predicting stable compounds and structures. A number of applications[19,29-33] illustrate its predictive power. In the structure prediction, we used a plane-wave kinetic energy cutoff of 650 eV, and the Brillouin zone was sampled with a resolution of $2\pi \times 0.05$ Å$^{-1}$, which showed excellent convergence of the energy differences, stress tensors and structural parameters. For further band structure calculation we used a higher energy cutoff of 850 eV. Phonon spectrum is calculated by the PHONOPY code[34]. The nodal- line structure was found with the aid of WannierTools[35].

**Acknowledgements**. This work was supported by the National Science Foundation of China (Grant No. 11674176), Tianjin Science Foundation for Distinguished Young Scholars (Grant No. 17JCJQJC44400), and the 111 Project (Grant B07013). Q.W. and O.V.Y. acknowledge support by the NCCR Marvel. The calculation was performed on the TianheII supercomputer at Chinese National Supercomputer Center in Guangzhou and at the Swiss National Supercomputing Centre (CSCS) under Project No. s675.

**Author Contributions** X. F. Z. and X. D. designed the research. X. D., X. F. Z., Q. W. and X. L. H. performed the calculations. All of authors analyzed the results, X. D., X. F. Z., O. V. Y. and Q.

W. wrote the paper.

**Author information** The authors declare no competing financial interests. Correspondence and requests for materials should be addressed to X. F. Z. (xfzhou@nankai.edu.cn) and H. T. W (htwang@nankai.edu.cn).